\documentclass [12pt]{article}

\hoffset = - 9 mm
\voffset = - 5 mm
\usepackage{graphicx}
\begin{document}
\thispagestyle{empty}
 \begin{center} {\Large\bf
On electromagnetic induction}\\
\vskip1truecm \small Giuseppe Giuliani\\ Dipartimento di
Fisica `Volta', Via Bassi 6, 27100 Pavia, Italy\\
\vskip0.25truecm
Email: giuliani@fisav.unipv.it\\
\vskip0.25truecm
Web site~-~http://matsci.unipv.it/percorsi/
 
 \end{center}
 
\vskip1truecm
\noindent {\bf Abstract}.
A general law for electromagnetic induction phenomena
is derived from Lorentz force and Maxwell equation connecting electric
field and time variation of
magnetic field. The derivation provides with a unified mathematical treatment
 the
statement according to which electromagnetic induction is
the product of two independent phenomena: time variation of magnetic
field and effects of magnetic field on moving charges.
The  general law  deals easily~-~without {\it ad hoc}
assumptions~-~with
typical cases usually considered as exceptions to the
flux rule and contains the  flux rule as a
particular case.
\tableofcontents
 \newpage
\section{Introduction}\label{intro} It is, in general, acknowledged that the
theoretical treatment of electromagnetic induction phenomena
presents some problems when part of the electrical circuit is
moving. Some authors speak of exceptions to
the flux rule;\footnote{\label{fey} R. Feynman,
R. Leighton and M. Sands {\it The Feynman
Lectures on Physics}, vol. II, (Addison Wesley, Reading, Ma.,  1964 ), pp.
17~-~2,3.} others save
 the flux rule by {\it ad hoc} choices
of the integration line  over which  the induced
$ emf$ is calculated. Several
attempts to overcome these difficulties have been made; a comprehensive one
 has been performed by Scanlon,
 Henriksen and Allen.\footnote{\label{69} P.J. Scanlon,
R.N. Henriksen and J.R. Allen, ``Approaches to electromagnetic induction'',
{ Am. J.  Phys.}, {37}, (1969), 698~-~708.}
 However,
their treatment~-~as others~-~fails to recognize that one must distinguish
between the velocity of the circuit elements and the velocity of the
electrical charges (see section \ref{velo} below). Therefore, these
authors reestablish the flux rule and, consequently, do not solve the
problems posed by its application.
\par
Since 1992, I have been teaching electromagnetism in a course  for Mathematics students
and I had to deal with  the problems outlined above.
 I have found that it is possible to  get
   a general
law for electromagnetic induction that contains the standard flux rule as
a particular case.\par
The matter has conceptual relevance;  it has also
historical and epistemological aspects that deserve to be investigated. Therefore,
it is, perhaps, worthwhile to submit the
following considerations to the attention of a public wider than that
of my students.
\section{\label{velo} A general law for electromagnetic induction}
Textbooks show a great variety of
positions about how the flux rule can be applied to the known
experimental phenomena of electromagnetic induction. Among the more   lucid
approaches, let us refer to the treatment given by
Feynman, Leighton and Sands in the {\em Feynman Lectures on Physics}.
They write:
\begin{quote}
 In general, the force per unit charge is $\vec F/q=\vec E + \vec v \times \vec B$. In
moving wires there is the force from the second term.
Also, there is an $\vec E$ field if there is somewhere a changing
magnetic field. They are two independent effects, but the emf around the
loop of wire is always equal to
the rate of change of magnetic flux
through it.\footnote{\label{fey2} R. Feynman,
R. Leighton and M. Sands {\it The Feynman
Lectures on Physics}, vol. II,  (Addison Wesley, Reading, Ma., 1964), p.
17~-~2.}
\end{quote}
This sentence is followed by a paragraph entitled
{\em Exceptions to the ``flux rule''}, where
the authors treat two cases~-~the Faraday disc and the `rocking plates'~-~both
characterized by the fact that there is a part of the circuit in which the
{\em material of the circuit} is changing.
As the authors put it, at the end
 of the discussion:\footnote{\label{fey3}  Ibidem,  p.
17~-~3.}
\begin{quote}
The `flux rule does not work in this case.
It must be applied to circuits in which the {\em material} of the circuit
remain the same. When the material of the circuit is changing, we must return to the basic laws.
The {\em correct} physics is always given by the two basic laws
\begin{eqnarray}
\vec F &=& e(\vec E + \vec v \times \vec B)\label{forza}\\
\nabla \times \vec E & = & - {{\partial \vec B}\over{\partial t}} \label{maxwell}
\end{eqnarray}
\end{quote}
\subsection{\label{definition} A definition of $emf$}
In order to try shedding some more light on the subject,
let us begin with the acknowledgement that the
expression of Lorentz force
\begin{equation}
 \vec F = q(\vec E +\vec v \times \vec B)
\end{equation}
not only gives meaning to the fields solutions of Maxwell equations
 when applied to point charges, but yields new
 predictions.\footnote{\label{lorentz} The fact that
 the expression of Lorentz force can be derived by considering
 an inertial frame in which the charge is at rest and by {\it
 assuming} that the force acting on it is simply given by
$\vec F' = q\vec E'$, does not change the matter.}
\par
The velocity appearing in the expression of Lorentz force
is the velocity of the charge: from now on, we shall use
the symbol $\vec v_{charge}$ for distinguishing the charge velocity  from
 the velocity $\vec v _{line}$ of the circuit element that contains
 the charge.  {\em This is a basic point of the present treatment}.
\par
Let us consider the integral of
 $(\vec E +\vec v_{charge} \times \vec B)$ over a closed loop:
\begin{equation}  \label{forzaem} {\cal E}= \oint
_{l}^{}{(\vec E + \vec v_{charge} \times \vec
B)\,\cdot\,\vec{dl}} = \oint _{l}^{}{\vec E\,\cdot\, \vec{dl}}+
\oint _{l}^{}{(\vec v_{charge} \times \vec B)\,\cdot\, \vec{dl}}
\end{equation}
This integral yields the work done by the electromagnetic field
on a unit positive point charge along the closed path considered.
 It presents itself as the {\em natural} definition of
the electromotive force, within the Maxwell~-~Lorentz theory: $emf= {\cal E}$.
\par Let us now calculate the value of
$  \cal E$ given by equation (\ref{forzaem}).
 The calculation
of the first integral appearing in the third member of equation
(\ref{forzaem}) yields: \begin{equation}  \label{z1} \oint
_{l}^{}{\vec E\,\cdot\, \vec{dl}} = \int_{S}^{} rot\, \vec E \,
\cdot \, \hat n \, dS = - \int_{S}^{} {{\partial \vec B} \over
{\partial t}} \, \cdot \, \hat n \, dS \end{equation} where $S$
 is any surface having the line $  l$ as contour and where we
have made use of Maxwell equation
(\ref{maxwell}).
 The calculation of the last integral of
equation (\ref{z1}) yields: \begin{equation}
\label{flussog} \oint_{l}^{} \vec E \, \cdot \, \vec dl = -
{{d}\over{dt}}\int_{S}^{} {\vec B \, \cdot \, \hat n \, dS} -
\oint_{l}^{} (\vec v_{line} \times \vec B)\, \cdot \, \vec dl
\end{equation} where $  \vec{v}_{line}$ is the velocity of
the circuit element $  \vec {dl}$.\footnote{\label{somm} A. Sommerfeld, {\it Lectures
 in Theoretical Physics}, vol. II,
(Academic Press, New York) 1950, pp. 130~-~132; ibidem, vol. III, p. 286.}
Notice that  equation (\ref{flussog}) is the result of a theorem of vectorial
calculus that is valid for any vector field $\vec G$, like the
magnetic field, for which $div\, \vec G=0$.
Therefore, we
get:
\begin{equation}
 \label{flussogg} {\cal E} =\left[ -
{{d}\over{dt}}\int_{S}^{} {\vec B \, \cdot \, \hat n \, dS} -
\oint_{l}^{} (\vec v_{line} \times \vec B)\, \cdot \, \vec dl\right] +
 \oint_{l}^{} (\vec v_{charge} \times \vec B)\, \cdot \, \vec {dl}
\end{equation}
This  equation says that: \begin{enumerate} \item
The induced {\it
emf} is, in general, given by three terms. \item The first two~-~grouped
under square brackets for underlining their
common mathematical and physical origin~-~come from the line integral $  \oint _{l}^{}{\vec E\,\cdot\,
\vec{dl}}$, whose value is controlled by Maxwell equation
(\ref{maxwell}) through equation (\ref{z1}). Accordingly,
 {\it their sum must be zero
when the magnetic field does not depend on time}. In this case, the law assumes the simple form:
\begin{equation}\label{simple}
{\cal E} =
\oint_{l}^{} (\vec v_{charge} \times \vec B)\, \cdot \, \vec {dl}
\end{equation}
 \item The third
term comes from the magnetic component of Lorentz force; we shall
see later  how this term may be different from
zero.
\item The flux rule is contained in the general law as a particular case.
\end{enumerate}
The general law (\ref{flussogg}) can be written also in terms
of the vector potential $\vec A$.  If  we put
\begin{equation}
\vec E = - grad\, \varphi -{{\partial \vec A}\over{\partial t}}
\end{equation}
in the first integral of the third member of
equation (\ref{forzaem}),
we get at once
\begin{equation}\label{flusso_A}
{\cal E} =  - \oint_{l}^{} {{\partial \vec A}\over{\partial t}} \cdot \vec dl+
\oint_{l}^{} (\vec v_{charge} \times \vec B)\, \cdot \, \vec {dl}
\end{equation}
since $\oint_{l}^{}{grad\,\varphi \cdot \vec dl}=0$.
When the magnetic
field does not depend on time ($\partial \vec B/\partial t=0$), we have:
\begin{equation}\label{potvett}
\oint_{l}^{}{\vec E \cdot \vec dl} =-\oint_{l}^{}{{{\partial \vec A}\over{\partial t}}\cdot \vec dl}
= -\int _{S}^{} {rot\, {{\partial \vec A}\over{\partial t}}\cdot\hat n\, dS}
=0
\end{equation}
since $rot\,( \partial \vec A/\partial t )=0$, because $\partial \vec B/\partial t=0$.
Therefore,   the only surviving
term in the expression of the induced $emf$ is coming from the magnetic component
 of Lorentz force and the law assumes the  form given by equation (\ref{simple}).
\subsection{$\vec A$ versus $\vec B$}
Traditionally,  textbooks  use equations containing the magnetic
field in dealing with  induction phenomena: the $emf$ in a loop
is partially (general law) or
totally (flux rule) dependent on a surface integral of $\vec B$. The laws are predictive,
 but they are not causal laws.  The values of $\vec B$ over the
 surface of integration cannot be causally related to the value of the $emf$ around
 the circuit because $\vec B$  acts at a distance (with an
 infinite propagation velocity): {\em in this case}, $\vec B$ is not a {\em good}
 field, since a {\em good} field, in Feynman's words, can be defined as
 \begin{quote}
 \dots a set of numbers we specify in such a way that what happens
{\em at a point} depends only on the numbers {\em at that point}.
We do not need to know any more about what's
going on at other places.\footnote{\label{good} R. Feynman,
R. Leighton and M. Sands {\it The Feynman
Lectures on Physics}, vol. II,  (Addison Wesley, Reading, Ma., 1964), p.
15~-~7. Feynman, uses the term `real field' instead of `good field'. Feynman
deals with this problem in discussing the Bohm~-~Aharanov effect. It is interesting
to see that an identical situation arises in classical electromagnetism.
}
\end{quote}
 On the contrary, the same equations written in terms of the vector potential, are
 causal laws  since they relate the value of the
 $emf$ around the circuit to the values of $-\partial \vec A/\partial t$  at the points of
 the loop: the vector potential is, in this case, a {\em good} field.
\section {\label{come}How the
law works}
\subsection {General features\label{general}}
Let us come back to the general law (\ref{flussogg} or \ref{flusso_A}).
The charge velocity  appearing in these equations contributes to build up,
through the factor $\vec v_{charge}\times\vec B$ the induced electromotive  field.
The charge velocity is given by:
\begin{equation}
\vec v_{charge}=
\vec v_{line}+\vec v_{drift}
\end{equation}
where $\vec v_{drift}$ is the drift velocity.
\par
Therefore,
  the general equation for electromagnetic induction
assumes the form:\nopagebreak
\begin{eqnarray}
 \label{flussogg''} {\cal E} &=&\left[ -
{{d}\over{dt}}\int_{S}^{} {\vec B \, \cdot \, \hat n \, dS} -
\oint_{l}^{} (\vec v_{line} \times \vec B)\, \cdot \, \vec dl\right] + \nonumber\\
&+& \oint_{l}^{} (\vec v_{line} \times \vec B)\, \cdot \, \vec {dl} +
 \oint_{l}^{} (\vec v_{drift} \times \vec B)\, \cdot \, \vec {dl}
\end{eqnarray}
or, in terms of the vector potential:
\begin{equation}\label{flusso_A'}
{\cal E} =  - \oint_{l}^{} {{\partial \vec A}\over{\partial t}} \cdot \vec dl+
\oint_{l}^{} (\vec v_{line} \times \vec B)\, \cdot \, \vec {dl}  +
\oint_{l}^{} (\vec v_{drift} \times \vec B)\, \cdot \, \vec {dl}
\end{equation}
When the circuit is made by a loop of
wire,  equation  (\ref{flussogg''}) reduces to equation
\begin{equation}
 \label{flussogg'} {\cal E} =\left[ -
{{d}\over{dt}}\int_{S}^{} {\vec B \, \cdot \, \hat n \, dS} -
\oint_{l}^{} (\vec v_{line} \times \vec B)\, \cdot \, \vec dl\right] +
 \oint_{l}^{} (\vec v_{charge/line} \times \vec B)\, \cdot \, \vec {dl}
\end{equation}
and equation (\ref{flusso_A'}) to equation
\begin{equation}\label{flusso_A''}
{\cal E} =  - \oint_{l}^{} {{\partial \vec A}\over{\partial t}} \cdot \vec dl+
\oint_{l}^{} (\vec v_{charge/line} \times \vec B)\, \cdot \, \vec {dl}
\end{equation}
because the drift
velocity is always parallel to the line element $\vec {dl}$ and, consequently,
 the integral
containing it is zero. We have introduced the new notation $v_{charge/line}$
for remembering that we are dealing with the velocity of the
charge that, in this case, can be replaced by
the velocity of the circuit element that contains it.
It is worth stressing again that, when the magnetic field does not depend on time,
equation (\ref{flussogg'}) cannot be read in terms of flux variation, since the
sum of the first two terms under square brackets  is zero as it is the equivalent
term of equation (\ref{flusso_A''}) containing the time derivative of the
vector potential.
 \par
When part of the circuit is made by extended material, the calculation of the
drift velocity contribution  to the induced $emf$ is not an easy task, since
the distribution of current lines in the
 material may be very complicated.\footnote{\label{hertz} The distribution of currents in extended materials have been studied
since the mid of eighteenth century. The following references are those that I know:
E. Jochmann, ``On the electric currents induced by a magnet in a rotating
conductor'', {\em Phil. Mag.}, {\bf 27}, (1864), 506~-~528;
{\em Phil. Mag.}, {\bf 28}, (1864), 347~-~349;
 H. Hertz, `On induction in rotating spheres', in {Miscellaneous
papers}, (J.A. Barth, Leipzig) 1895, pp. 35~-~126. A discussion
of some aspects of Hertz's work can be found in: Jed Z. Buchwald,
{The creation of scientific effects~-~Heinrich Hertz and electric waves},
(The University of Chicago Press, Chicago and London) 1994, pp. 95~-~103;
the papers by Boltzmann and Corbino quoted in footnotes \ref{corb};
\ref{boltz}. V. Volterra,   ``Sulle correnti elettriche in una lamina
metallica sotto l'azione di un campo magnetico'', Il Nuovo Cimento, {\bf 9}, 23~-~79
(1915). One should also see the literature on eddy currents.
}
In section \ref{corbino_sec} and \ref{far2} we shall treat  particularly simple cases
  with a circular symmetry.
In all other cases, we shall neglect the
drift velocity in the calculation of the induced $emf$ (look at the end
of section \ref{corbino_sec} for a further discussion of this point).
\subsection{How it works in specific cases}
We shall now discuss some cases widely treated in literature,
in order to see how the general law (and the flux rule) can be applied to
 specific problems.
\subsubsection {When a bar is moving}
\begin{figure}[htb]
 \centerline{
 \includegraphics{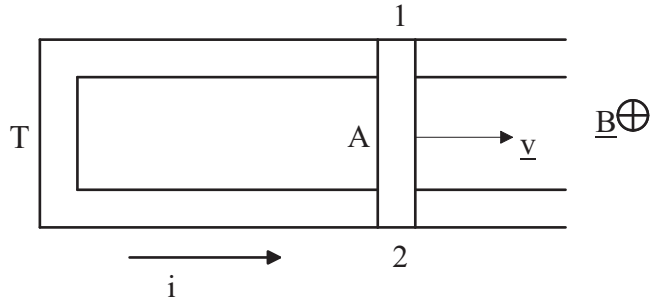}
 }
 \caption{  the conducting bar $A$, of length $a$, moves with constant
velocity $\vec v$ in a uniform and
constant magnetic field $\vec B$ perpendicular to and entering the page.
The bar $A$ slides on the conducting frame $T$. A steady current $i$
circulates in the circuit  $T21$.
}
 \end{figure}
Let us consider the circuit of fig. 1. The conducting bar $
A$, of length $a$, slides with constant velocity $\vec v$ over the  conducting frame $  T$
in a uniform and constant magnetic field $  \vec B$ perpendicular
to the plane of the frame and entering the page: the frame $T$ is at rest
with respect to the source of the magnetic field. A steady current flows
along the direction indicated in the figure.
Notice that in this case we can neglect without approximation the
drift velocity of the charge because it is always directed along the integration line
(owing to the Hall effect).
 According to the
general law (\ref{flusso_A''}), the first integral is zero as zero is
sum of the first two terms of equation (\ref{flussogg'}):
 as a matter of fact, if we choose the counterclockwise direction for the line integral,
 we obtain for the sum of these two terms:
 \begin{equation}
 Ba v_{line} - Ba v_{line} =0
\end{equation}
The induced electromotive force is then given by:
\begin{equation}\label{ebarra}
{\cal E} =Ba v_{charge/line}
\end{equation}
The surviving term {is} the one
coming from the
magnetic component of Lorentz force.
 Moreover, the theory predicts that the {\it emf} is
localized into the bar: the bar acts as a battery and, as a consequence,
 between the two points
$  1$ and $  2$ of the frame that are in contact with
the bar, one should measure a potential difference given by $
 V_1 -V_2 ={\cal E}-ir$, where $  i$ is the circulating
current and $  r$ is the resistance of the bar.
\par
Finally, it is worth stressing that the energy balance
 shows that the magnetic component of
Lorentz force plays the role of an intermediary: the electrical
energy dissipated in the circuit comes from the work done by the
force that  must be applied  to keep the bar moving with
constant velocity.
 \par Let us now recall
how the  flux rule deals with this case. It predicts an
{\it emf} given by $  Bav_{line}$. In the light of the general law (\ref{flussogg'})
and of its discussion,
we  understand why the flux rule predicts correctly the value
of the {\it emf}: the reason lies in the fact that the two line
integrals  under square brackets cancel each
other. However, we have shown above that the physics embedded in the general
equation (\ref{flussogg'})
forbids to read equation (\ref{ebarra}) as the result of:
\begin{equation}\label{ebarran}
{\cal E}= Ba v_{line} + [-Bav_{line} + Ba v_{line}]  =Ba v_{line}
\end{equation}
 that leaves operative the first term coming from the flux variation.
\par
Before leaving this subject, it is worth saying something about
the long debate about the meaning of `localized $emf$'.
With reference to the moving bar, we can proceed as follows.
The points $1$ and $2$ divide the circuit into two parts: the one on the left
 obeys Ohm law $\Delta V = iR$ and
the current enters it at point $1$  at higher potential  and leaves
it  at point $2$ at
lower potential; the part on the right does not obeys Ohm law and
the current enters it at point $2$  at lower potential  and leaves
it  at point $1$ at
higher potential. We say that the $emf$ is localized in the right part
of the circuit: it can be experimentally distinguished from the other.
 
\subsubsection{The Faraday disc: 1\label{faraday_disc}}
 \begin{figure}[htb]
 \centerline{
 \includegraphics{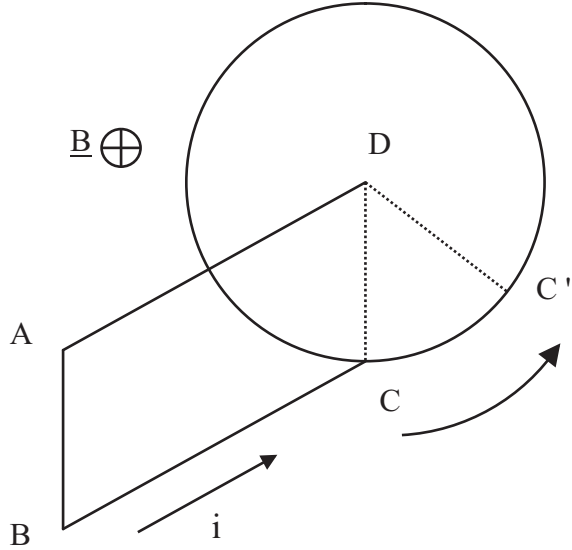}
 }
 \caption{   the Faraday disc. A conducting disc rotates with constant
angular velocity $\omega$
in a uniform and constant magnetic field $\vec B$ perpendicular to and
entering the plane of the disc. A conducting frame $DABC$ makes conducting
contacts with the center and a point on the periphery of the disc.
A steady current $i$
circulates as indicated by the arrow.
}
 \end{figure}
A conducting disc is rotating with
constant angular velocity $  \omega$ in a uniform and
constant magnetic field $   \vec B$ perpendicular to the disc and
entering the page (fig. 2).  A conducting frame $DABC$ makes conducting
contacts with the center and a point on the periphery of the disc.
A steady current $i$
circulates as indicated by the arrow.
\par
By applying the general laws  (\ref{flusso_A''}) or  (\ref{flussogg'}) to the integration
line at rest $  ABCDA$, we see that, if we neglect the contribution
of the drift velocity:
\begin{itemize}
\item since the magnetic field does not depend on time,
 the first integral in (\ref{flusso_A''}) is zero
\item equivalently,  the sum
of the first two integrals in (\ref{flussogg'}) is zero.
In the present case, each of the two terms is zero: the flux
associated with the circuit does not change and $  \vec
v_{line}$ is zero everywhere, since the integration line has been chosen at rest
\item the only surviving term is the one due to the magnetic
 component of Lorentz force and its value is given by
  (we are taking the counterclockwise direction
for the line integration): \begin{equation}  \label{disco}
{\cal E} = {{1}\over{2}}\omega B R^2 \end{equation} where $
R$ is the disc radius. In doing this calculation, we have neglected~-~as explained
at the end of section \ref{general},
the charge drift velocity (we shall take it into account in section \ref{far2}).
\end{itemize}
\noindent
 Of course, the {\it emf}
given by equation (\ref{disco}) is induced along any radius, as
it can be seen by considering the circuit $  ABCC'DA$. If the radius $C'D$
is considered at rest, the case is the same as that of the circuit $ABCDA$ discussed just
before. If
the radius $  C'D$ is considered in motion with angular velocity $ \omega$, we see
 again,  both from equation  (\ref{flusso_A''}) and equation (\ref{flussogg'}), that
 the only surviving term is the one due to the magnetic component
 of Lorentz force with the only difference that, now,
 the  first term of  equation (\ref{flussogg'})
$  ({{1}/{2}})\omega B R^2$ is
cancelled out by the second $  -({{1}/{2}})\omega B R^2$, due
to the movement of the line elements of the radius $  DC'$;  the line elements lying on the
$  CC'$ arc give a null contribution.
 \par
Let us now see how  the flux rule deals with the Faraday disc.
If we choose the lines {\em at rest} $ABCDA$ or $  ABCC'DA$,
 we find  that the induced {\it emf}
is zero;  if we choose the integration line $ABCC'DA$, with the radius
$C'D$ {\em considered in motion},
we obtain the correct  result
${\cal E}=  ({{1}/{2}})\omega B R^2$.
\par
 As in the case
 of the moving bar, the flux rule gets
  the correct result
 only because the two line integrals under square brackets in  equation  (\ref{flussogg'})
  cancel each other.
 As in the case of the moving bar, the physics embedded in the general equation
 forbids an interpretation of the mathematical result in terms of flux variation:
 again the physical origin of the induced {\it emf} is due to the
 intermediacy of the magnetic component
 of  Lorentz force.
  \subsubsection{The unipolar induction}
  \begin{figure}[htb]
 \centerline{
 \includegraphics{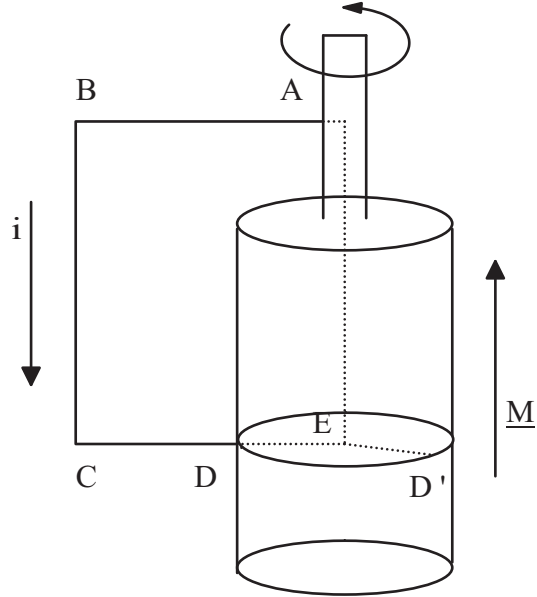}
 }
 \caption{    the unipolar induction. A conducting cylindrical magnet rotates about its
axis in a counterclockwise direction. A conducting frame $ABCD$ makes
conducting contacts between a point on the axis of the magnet and
a point on its surface. A steady current $i$
circulates as indicated by the arrow.
}
 \end{figure}
The
so called unipolar induction is illustrated in fig. 3.
When the cylindrical and conducting magnet $  M$ rotates
about its axis with angular velocity $  \omega$ in the
counterclockwise direction, a current flows in the circuit
 as indicated by the arrow. It is easy to see that the
discussion of the unipolar induction can be reduced to that of
the Faraday disc for both  general law and flux rule: the
general law applied to the circuit $  ABCDEA$ at rest yields an
$emf$ induced  in the radius $  DE$ given by  $
(1/2)\omega B \overline{DE}^2$, while the flux rule applied to
the same circuit predicts a zero {\it emf}. If  one
consider instead the integration line  $ABCDD'EA$ with the radius $D'E$ in motion,
analogous to the path $ABCC'DA$
of fig. 2 concerning the Faraday disc,
 one can follow the same arguments
developed in that case.
\subsubsection{The flux varies, but\dots}
\begin{figure}[htb]
 \centerline{
 \includegraphics{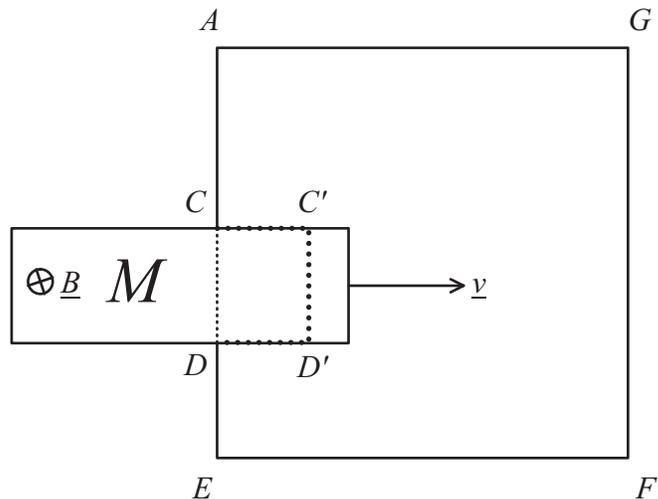} }
 \caption{    Kaempffer's example. The circuit $DEFGAC$ is closed by the
conducting magnet $M$ that is moving to the right
with constant velocity $\vec v$. The magnet is sufficiently long in the
direction perpendicular to the page so that there is a magnetic
field only within the magnet. There is no induced $emf$ in spite of the
fact that there is a flux change through the circuit at rest $ACDEFGA$.
}
 \end{figure}
In fig. 4 a case discussed by Scanlon,
 Henriksen and  Allen\footnote{\label{scan2}  P.J. Scanlon,
R.N. Henriksen and J.R. Allen, ``Approaches to electromagnetic induction'',
{ Am. Jour.  Phys.}, {\bf 37}, pp. 705~-~706 (1969).}
and
originally due to Kaempffer\footnote{\label{k} F. A. Kaempffer, {\em Elements of Physics} (Blaisdell Publ. Co., Waltham,
Mass., 1967), p. 164; quoted by   Scanlon,
 Henriksen and  Allen.} is presented.
When the conducting magnet $M$ moves, as indicated, with constant velocity $\vec v$,
there is no induced $emf$~-~if the magnet is sufficiently long in the
direction perpendicular to the page so that there is a magnetic
field only within the magnet. If we consider the integration line $ACDEFGA$ at rest,
there is a flux variation without induced $emf$. Also in this case,
we must choose {\em ad hoc} the integration line~-~for instance the line
$ACC'D'DEFGA$, where the segment $C'D'$ moves with the magnet~-~if we want
to save the flux rule. On the other hand, the general law works well,
whatever integration line is chosen.
\subsubsection{The `rocking plates'}
\begin{figure}[htb]
 \centerline{
 \includegraphics{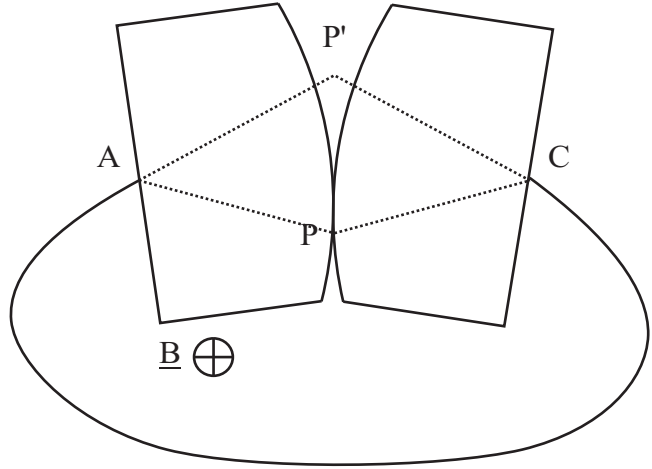} }
 \caption{   the `rocking plates'. Two conducting plates  oscillate back and forth
in a uniform and constant magnetic field perpendicular to the plates.
}
 \end{figure}
This is one of the two `exceptions to the flux rule'  discussed by Feynman,  Leighton and
Sands (fig. 5).\footnote{\label{rocking} R. Feynman,
R. Leighton and M. Sands {\it The Feynman
Lectures on Physics}, (Addison Wesley, Reading, Ma.,  1964 ), pp.
17~-~3.}
The two plates oscillate slowly back and forth so
that their point of contact moves from $P$ to $P'$ and viceversa. The circuit is
closed by a wire that connects point $A$ and $C$. The magnetic field $\vec B$
is perpendicular to the plates and enters the page. The authors write:
\begin{quote}
If we imagine the ``circuit'' to be completed through the plates on the dotted line
shown in the figure, the magnetic flux through this circuit changes by
a large amount as the plates are rocked back and forth. Yet the rocking can be done
with small motions, so that $\vec v \times \vec B$ is very small and there is
practically  no emf.
\end{quote}
According to the general law (\ref{flussogg'})
the sum of the first two terms of equation (\ref{flussogg'}) must be zero. Hence:
\begin{equation}
 \oint_{l}^{} (\vec v_{line} \times \vec B)\, \cdot \, \vec dl = -
{{d}\over{dt}}\int_{S}^{} {\vec B \, \cdot \, \hat n \, dS}
\end{equation}
Therefore, the third term of equation (\ref{flussogg'}) coming from
 the magnetic component of Lorentz force equals
the magnetic flux variation changed in sign, if, of course, we assume
 that the charge velocity  can be taken equal to the
velocity of the element of conductor that contains it (the drift velocity
is neglected). The conclusion is
that there is an induced $emf$: its average value is given by $-\Delta \phi/
\Delta t$ where $\Delta \phi$ is the flux variation between the two
extreme positions of the plates and $\Delta t$ the interval of time
taken in going from one position to the other: the induced $emf$ gets
smaller and smaller as $\Delta t$ gets larger and larger; when the motion of the plates
is very slow, we can conclude with Feynman that `there is
practically  no $emf$'.  Notice that the
induced $emf$ changes in sign (and the current its direction) when the the rocking
motion is reversed.
\subsubsection{The Corbino disc\label{corbino_sec}}
The discussion of this case will show how the charge drift velocity,
always neglected before, plays its role in the building up of the
induced electromotive field and, {\em therefore}, of the induced $emf$.
In 1911, Corbino studied theoretically and experimentally the
case of a conducting
disc with a hole at its center.
If a voltage is applied between the inner and the outer periphery of the disc, a
radial current will flow, provided that the experimental setup is
realized in a way suitable for maintaining the circular symmetry:
the inner and outer periphery are covered by highly conducting electrodes; therefore,
the inner and outer periphery are two equipotential lines.
If a uniform and constant magnetic field is applied perpendicularly
to the disc, a circular
 current will flow in the disc.\footnote{\label{corb} O.M.
Corbino,    ``Azioni elettromagnetiche dovute
agli ioni dei metalli deviati dalla traiettoria
normale per effetto di un campo'', { Il Nuovo Cimento} {\bf 1}, 397~-~419  (1911).
A german translation of this paper appeared in Phys. Zeits., {\bf 12},
561~-~568 (1911). For a historical reconstruction see: S. Galdabini and G.
Giuliani, ``Magnetic field effects and dualistic theory of metallic
conduction in Italy (1911~-~1926): cultural heritage, epistemological beliefs,
and national scientific community'', Ann. Science {\bf 48}, 21~-~37 (1991).
As pointed out by von Klitzing, the quantum Hall effect may be considered as
an ideal (and quantized)
version of the Corbino effect
corresponding to the case in which the current in the disc, with an
applied radial voltage, is only circular: K. von Klitzing,
``The ideal Corbino effect'', in: P.E. Giua ed., {\em Commemorazione di Orso Mario Corbino},
(Centro Stampa De Vittoria, Roma, 1987), pp. 43~-~58.
}
\par
The first theoretical treatment of this case is due, as far as I know,
to Boltzmann who wrote down the equations of motion of charges
in combined electric
 and magnetic fields.\footnote{\label{boltz} L. Boltzmann, { Anzeiger der Kaiserlichen Akademie der
Wissenschaften in Wien},{\bf 23}, (1886), 77~-~80; {Phil. Mag.}, {\bf 22},
(1886), 226~-~228.}
 Corbino, apparently not aware of this fact,
obtained the same equations already developed by Boltzmann. However,
while Boltzmann focused on magnetoresistance effects, Corbino
interpreted the theoretical results in terms of radial and circular currents
and studied experimentally the magnetic effects due to the latter ones.
\par
The application of the general law of electromagnetic
induction to this case
leads to the same results usually obtained (as Boltzmann and Corbino did)
by writing down and solving the equations of motion of the charges in an
electromagnetic field (by taking into account, explicitly or
implicitly, the scattering processes).
\par
If $I_{radial}$ is the radial current, the radial current density $J(r)$ will be:
\begin{equation}
 J(r) = {{I_{radial}}\over{2\pi r s}}
\end{equation}
and the radial drift velocity:
\begin{equation}\label{radial_d}
 v(r)_{drift}= {{I_{radial}}\over{2\pi r s n e}}
\end{equation}
where $s$ is the thickness of the disc, $n$ the electron concentration
and $e$ the electron charge.
In the present case the general law of electromagnetic induction assumes
the simple form of equation (\ref{simple}) with $v_{charge}=v_{drift}$;
therefore,
the induced $emf$ around a circle of radius $r$ is given by:
\begin{equation}
{\cal E}_{circular}= \oint_{0}^{2\pi r}{(\vec v(r)_{drift}\times \vec B})\cdot \vec{dr} =
{{I_{radial}\, B}\over{s ne}}
\end{equation}
The circular current $dI(r)_{circular}$ flowing in a circular  strip
of radius $r$ and section $s\cdot dr$ will be, if $\rho$ is the resistivity:
\begin{equation}\label{corre_circ}
 dI_{circular} =   {{{\cal E}_{circular}\, s dr}\over{ \rho\, 2\pi r}}= {{\mu B}\over{2\pi}} I_{radial}  {{{dr}\over{r}}}
\end{equation}
and the total circular current:
\begin{equation}\label{circolare}
I_{circular} =  {{\mu B}\over{2\pi}}  I_{radial} \ln {{r_2}\over{r_1}}
\end{equation}
where $\mu$ is the electron mobility, $r_1$ and $r_2$ the inner and outer radius
of the disc (we have used the relation $\mu = 1/\rho n e$).
Equation (\ref{circolare}) is the same as that derived and experimentally
tested by Corbino.
\par
The power dissipated in the disc is:
\begin{equation} \label{power}
 W= (I^2 R)_{radial} + (I^2 R)_{circular}=
  I_{radial}^2 R_{radial}(1+ \mu^2 B^2)
\end{equation}
where we have used equation (\ref{circolare}) and the  two  relations:
\begin{eqnarray}
R_{radial} & = & {{\rho}\over{2\pi s}} \ln {{r_2}\over{r_1}}\\
R_{circular} & = & {{\rho^2}\over{s^2}}{{1}\over{R_{radial}}}
\end{eqnarray}
Equation (\ref{power}) shows that the phenomenon may be described as due to an increased
resistance $R_{radial}(1+ \mu^2 B^2)  $: this is the magnetoresistance effect.
The circular induced $emf$ is `distributed' homogeneously along each circle.
Each circular strip of section $s\cdot dr$ acts as a battery that produces
current in its own resistance: therefore, the potential difference
between two points arbitrarily chosen on a circle is zero.
 Hence, as it must be, each circle
is an equipotential line.
\par
The above application  of the
general law yields  a  description  of Corbino disc that combine Boltzmann
(magnetoresistance effects) and Corbino (circular currents) point of view
 and shows how the general law
can be applied to phenomena traditionally considered outside the phenomenological domain
for which it has been derived.
\subsubsection{The Faraday disc: 2\label{far2}}
The discussion of Corbino disc helps us in better understanding the physics
of the Faraday disc.
Let us consider a Faraday disc in which the circular symmetry is conserved.
This may be difficult to realize; anyway,  it is interesting to
discuss it.
As shown above, the steady condition will be characterized by the flow of a radial
and of a circular current.
The mechanical power needed to keep the disc rotating with constant
angular velocity $\omega$ is equal to the work per unit time done by the
magnetic field on the rotating radial currents. Then, it will be given by:
\begin{equation}
W = \int _{0}^{2\pi} {}\int _{r_1}^{r_2} ({J_{radial}\,r\,d\alpha}\,s)(B\,dr)(\omega \, r)=
I_{radial}\,{{1}\over{2}}\,\omega\,B\, (r_2^2-r_1^2)
\end{equation}
where the symbols are the same as those used in the previous section.
The point is that the term
\begin{equation}
{\cal E}= {{1}\over{2}}\,\omega\,B\, (r_2^2-r_1^2)
\end{equation}
is the induced $emf$ due only to the motion of the disc. This $emf$ is the primary
source of the induced currents, radial and circular.
Therefore, the physics of the Faraday disc with circular symmetry,
may be summarized as follows:
\begin{enumerate}
\item [a)]  the source of the induced currents is the induced $emf$ due
to the rotation of the disc
\item [b)] the primary product of the induced $emf$ is a radial  current
\item [c)] the drift velocity of the radial current produces in turn a circular
induced $emf$ that give rise to the circular current
\end{enumerate}
 
\section{General law and flux rule}
The cases discussed above must be considered as illustrative examples.
Obviously,
the conceptual foundations of the general law and of the flux rule
can be discussed without any reference to particular cases.
The general law has been already dealt with in great detail.
Let us now focus our attention on the flux rule.
\par
Let us define~-~this, of course, is not our choice~-~the induced {\it emf}  as
\begin{equation}\label{standard}
 {\cal E} = \oint_l \vec E \cdot \vec dl
\end{equation}
where $\vec E$ obeys Maxwell equations.
  As shown above (equations
\ref{z1}~-~\ref{flussog}), this leads immediately to the conclusion that the
induced  {\it emf} is given by:
\begin{equation}
\label{flussog_2} {\cal E}=\oint_{l}^{} \vec E \, \cdot \, \vec dl =
- {{d}\over{dt}}\int_{S}^{} {\vec B \, \cdot \, \hat n \, dS} -
\oint_{l}^{} (\vec v_{line} \times \vec B)\, \cdot \, \vec dl
\end{equation}
or, in terms  of the vector potential, by (equations \ref{flusso_A}~-~\ref{potvett}):
\begin{equation}
{\cal E}=\oint_{l}^{}{\vec E \cdot \vec dl} =-\oint_{l}^{}{{{\partial \vec A}\over{\partial t}}\cdot \vec dl}
\end{equation}
and is always zero when $\partial \vec B/
\partial t=0$.
\par
 Of course, this result is not satisfactory, if we want
to save the flux rule. Therefore, one may
suggest to define the induced {\it emf} as:
\begin{equation}\label{allen}
 {\cal E}=\oint_{l}^{} \vec E \, \cdot \, \vec dl  +
\oint_{l}^{} (\vec v_{line} \times \vec B)\, \cdot \, \vec dl=
 -
{{d}\over{dt}}\int_{S}^{} {\vec B \, \cdot \, \hat n \, dS}
\end{equation}
thus reestablishing the flux rule. This is, for instance, the choice made by
Scanlon, Henriksen and  Allen.\footnote{\label{scan3} P.J. Scanlon,
R.N. Henriksen and J.R. Allen, ``Approaches to electromagnetic induction'',
{ Am. Jour.  Phys.}, {\bf 37},  (1969), p. 701.
} However:
\begin{itemize}
\item   since the velocity appearing in
 equation (\ref{allen}) is clearly the velocity
of the line element $\vec dl$, this  definition
has no physical meaning.
\item   the definition  is
  {\em ad hoc}. As a consequence, as in the paper by
Scanlon, Henriksen and  Allen, the line of integration must be chosen {\em ad hoc}
for obtaining predictions in accordance with the experimental findings.
\end{itemize}
\section{Conclusions} The straightforward
application of two fundamental laws of
electromagnetism~--~Maxwell equation $  rot \vec E =
-\partial \vec B/\partial t$ and Lorentz law $  \vec F = q(\vec E + \vec
v \times \vec B)$~--~leads to a general law for
electromagnetic induction phenomena: it includes, as a particular case,
the standard flux rule. The treatment given in this paper
shows that standard derivations:
\begin{enumerate}
\item [a)]  Fail to recognize that the
velocity involved in the integration of Maxwell equation
(\ref{maxwell}) is the velocity of the line elements composing the
closed path of integration and {\it not} the velocity of the
charges.
\item [ b)]  Overlook the fact that the expression of Lorentz
force constitutes a fundamental postulate not included in Maxwell
equations; this postulate not only gives physical meaning to the
fields solutions of Maxwell equations~--~when applied to
charges~--~but allows also the prediction of new phenomena.
\item [ c)] In relevant cases,
 when a part of the circuit is moving, they are misleading as far as the
  physical
origin of the phenomenon is concerned: they attribute the effect to
 a flux variation when, instead, the origin lies in the intermediacy of the magnetic
 component of Lorentz force.
\end{enumerate}
Finally, the present derivation:
\begin{itemize}
\item provides with a rigorous mathematical treatment
 the
statement according to which electromagnetic induction phenomena
are the product of two independent processes: time variation of magnetic
field and effects of magnetic field on moving charges.
\item yields a law that can be successfully applied to
 phenomena~-~those presented by Corbino disc~-~till now considered outside the phenomenological
 domain of electromagnetic induction.
 
\end{itemize}

 \par\vskip1cm\noindent
{\bf Acknowledgements}. The author would like to thank Ilaria
 Bonizzoni for a critical reading of the manuscript and
 Giuliano Bellodi  and Gianluca Introzzi for valuable discussions.
\end{document}